\documentclass[%
 aip,
 amsmath,amssymb,
 reprint,%
floatfix
]{revtex4-1}

\usepackage{graphicx}
\usepackage{dcolumn}
\usepackage{bm}

\usepackage[utf8]{inputenc}
\usepackage[T1]{fontenc}
\usepackage{mathptmx}
\usepackage{etoolbox}
\usepackage{hyperref}
\usepackage{booktabs}
\usepackage{multirow}
\usepackage{makecell}

\makeatletter
\def\@email#1#2{%
 \endgroup
 \patchcmd{\titleblock@produce}
  {\frontmatter@RRAPformat}
  {\frontmatter@RRAPformat{\produce@RRAP{*#1\href{mailto:#2}{#2}}}\frontmatter@RRAPformat}
  {}{}
}%
\makeatother
\begin{document}

\preprint{AIP/123-QED}

\title[XBTorch: A Unified Framework for Modeling and Co-Design of Crossbar-Based Deep Learning Accelerators]{XBTorch: A Unified Framework for Modeling and Co-Design of Crossbar-Based Deep Learning Accelerators}
\author{Osama Yousuf}
\affiliation{ 
George Washington University
}%
\affiliation{ 
DEVCOM Army Research Laboratory
}%
\affiliation{ 
Western Digital Research
}%

\author{Andreu L. Glasmann}%
\affiliation{ 
DEVCOM Army Research Laboratory
}%

 \author{Martin Lueker-Boden}%
\affiliation{ 
Western Digital Research
}%

 \author{Sina Najmaei}%
\affiliation{ 
DEVCOM Army Research Laboratory
}%

 \author{Gina C. Adam*}%
 \email{GinaAdam@gwu.edu}
\affiliation{ 
George Washington University
}%

\date{\today}

\begin{abstract}
Emerging memory technologies have gained significant attention as a promising pathway to overcome the limitations of conventional computing architectures in deep learning applications. By enabling computation directly within memory, these technologies - built on nanoscale devices with tunable and nonvolatile conductance - offer the potential to drastically reduce energy consumption and latency compared to traditional von Neumann systems. This paper introduces XBTorch (short for CrossBarTorch), a novel simulation framework that integrates seamlessly with PyTorch and provides specialized tools for accurately and efficiently modeling crossbar-based systems based on emerging memory technologies. Through detailed comparisons and case studies involving hardware-aware training and inference, we demonstrate how XBTorch offers a unified interface for key research areas such as device-level modeling, cross-layer co-design, and inference-time fault tolerance. While exemplar studies utilize ferroelectric field-effect transistor (FeFET) models, the framework remains technology-agnostic - supporting other emerging memories such as resistive RAM (ReRAM), as well as enabling user-defined custom device models. The code is publicly available at: \url{https://github.com/ADAM-Lab-GW/xbtorch}. 
\end{abstract}

\maketitle

\section{Introduction}

The exponential growth in artificial intelligence applications has driven the development of increasingly complex neural network architectures with growing computational demands. Traditional computing systems based on the von Neumann architecture face fundamental limitations in energy efficiency and performance due to the well-known ``memory wall,'' where data transfer between processing and memory units becomes the primary bottleneck \citep{zou2021breaking}. This
limitation has spurred interest in new computing paradigms that can overcome these constraints. 

One such paradigm that has gained significant attention is in-memory computing \citep{yu2025full, huang2020memory}. These systems  seek to overcome the limitations of traditional von Neumann architectures by performing computations directly within memory, reducing data movement and improving efficiency. A key enabler of in-memory computing is memristors, which have emerged as promising devices due to their non-volatility, low power consumption, and history-dependent conductance \citep{sun2021future}. These properties make memristors well-suited for accelerating neural network operations, particularly multiply-accumulate (MAC) computations, which are fundamental to deep learning. By performing MACs directly within memory, memristor-based systems can reduce energy consumption by up to five orders of magnitude compared to conventional CPUs \citep{taha2013exploring}.

Various emerging memory technologies have demonstrated potential for neuromorphic computing, including resistive random-access memory (ReRAM) \citep{yu2018neuro}, magnetic tunnel junctions (MTJs) \citep{grollier2016spintronic}, ferroelectric field-effect transistors (FeFETs) \citep{jerry2017ferroelectric}, and phase-change memory (PCM) \citep{sebastian2019computational}. While each of these technologies presents unique advantages, such as energy efficiency, scalability, or retention characteristics, no single memristive device has emerged as the definitive choice for neuromorphic hardware. Determining the optimal technology requires further investigation into factors such as endurance, variability, and fabrication scalability \citep{burr2017neuromorphic}.

Regardless of the underlying device technology, memristive neural networks are inherently susceptible to noise and non-idealities due to device physics, leading to performance degradation compared to fully digital implementations. These challenges manifest in both training and inference workloads, necessitating advancements at the algorithmic level. Existing literature on algorithmic investigations in memristive neural networks can be broadly classified into three categories. The first focuses on device modeling, where researchers develop efficient and accurate simulation techniques for large-scale memristor arrays to capture device-level imperfections and their impact on neural network behavior \citep{peng2019dnn+, jiang2016compact, chatterjee2023ferroelectric}. The second category explores hardware-aware training methods, which incorporate the non-ideal characteristics of memristive devices into the training process to enhance robustness and adaptability \citep{gokmen2020algorithm, borders2024measurement, cai2020power, yousuf2025robust}. The third area addresses inference-time fault tolerance, aiming to mitigate performance degradation caused by device variability and system-level noise through error correction techniques and adaptive computing strategies \citep{yousuf2025layer, joksas2020committee, huangfu2017computation}.

However, existing research in these areas is often technology-specific or scattered, lacking generalizable frameworks that accommodate diverse memristor technologies and their unique characteristics. This fragmentation hinders comparative studies and slows progress toward robust, scalable memristive neural networks. To address this gap, we introduce XBTorch, a modular simulation framework designed to facilitate algorithmic research across different memristive technologies. XBTorch seamlessly integrates with PyTorch, enabling users to conduct investigations in device modeling, hardware-aware training, and fault tolerance without requiring extensive modifications to their existing workflows. The core library is implemented entirely in Python and built on top of PyTorch, reusing its native vectorized operations and CUDA acceleration to maintain compatibility with standard ML workflows and minimize additional overhead during simulation and training. 

\section{Related Work}

There are numerous crossbar-based deep neural network simulators for emerging memory technologies that offer important functionality for simulating training as well as inference workloads. A few that are close to our work are NeuroSim \citep{peng2019dnn+}, IBM's Analog Hardware Acceleration Kit (AIHWKit) \citep{rasch2021flexible}, and CrossSim \citep{CrossSim}. We offer brief comparisons of each of these against XBTorch in the following sections.

\subsection{NeuroSim}
NeuroSim excels in system-level performance estimation - providing detailed modeling of energy consumption, computational latency, and area efficiency. However, it lacks support for hardware-aware training and fault-tolerance techniques that are essential for improving the robustness of emerging-memory-based neural systems. XBTorch, on the other hand, implements these advanced capabilities, integrating both existing and novel methodologies for hardware-aware training and inference. This design choice allows researchers to explore algorithm–hardware co-optimization strategies within a unified deep learning framework. However, XBTorch intentionally trades detailed system-level performance estimation for these enhanced modeling and co-design features, focusing instead on the algorithmic and device-interaction aspects of emerging memory technologies.

\subsection{AIHWKit}
IBM AIHWKit stands out for its comprehensive support across multiple aspects of analog in-memory computing. However, it does not provide built-in implementations of fault-tolerance mechanisms, requiring substantial modifications to the codebase for researchers interested in investigating these techniques. This challenge is further exacerbated by its reliance on a CUDA backend, which, while beneficial for accelerating large-scale workloads, imposes restrictions on customizability and necessitates expertise in GPU programming. In contrast, XBTorch maintains a high-level Python-based architecture, ensuring that algorithmic modifications remain accessible and straightforward. Additionally, XBTorch natively incorporates state-of-the-art fault-tolerance techniques from literature, facilitating comparative studies without requiring extensive modifications. Notably, a recent derivative, AIHWKit-lightning \citep{buchel2024aihwkit}, introduces simplifications that improve hardware-aware training efficiency, which is particularly valuable for simulating large-scale models with billions of parameters. For instance, it optimizes weight noise injection by applying it exclusively during the forward pass, which makes training large language models an order of magnitude faster than the base AIHWKIT simulator. Because of these faster training times, AIHWKIT-lightning would be the better candidate compared to XBTorch for large-scale hardware-aware training studies. We plan to integrate with optimized libraries for training and serving large models as part of a future release.

\subsection{CrossSim}
CrossSim distinguishes itself by allowing analog cores to function as drop-in replacements for NumPy arrays, enabling seamless emulation of analog hardware deployment beyond deep learning applications, such as general signal processing and solving linear systems. However, a notable limitation is that it only supports fully connected and convolutional neural network layers, restricting its applicability in modern deep learning architectures. In contrast, XBTorch is explicitly designed for neural network applications, leveraging PyTorch’s modularity. It wraps PyTorch layers while treating individual weights as non-ideal memristor devices, enabling seamless modeling of non-idealities during both inference and backpropagation, regardless of the underlying layer type.

In addition to the key differentiating factors described above, XBTorch provides several capabilities that further enhance its utility. It supports arbitrary quantization of network quantities, facilitating research into custom number formats for weights, activations, gradients, and error signals. Furthermore, it includes hardware-calibrated presets for various memristive devices and systems, as well as dedicated modules for evaluating the fidelity of different device models. 

The following section will outline XBTorch’s design philosophy and feature-specific implementations in greater detail.


\section{XBTorch}

\begin{figure*}[t]
  \centering
  \includegraphics[width=1\linewidth]{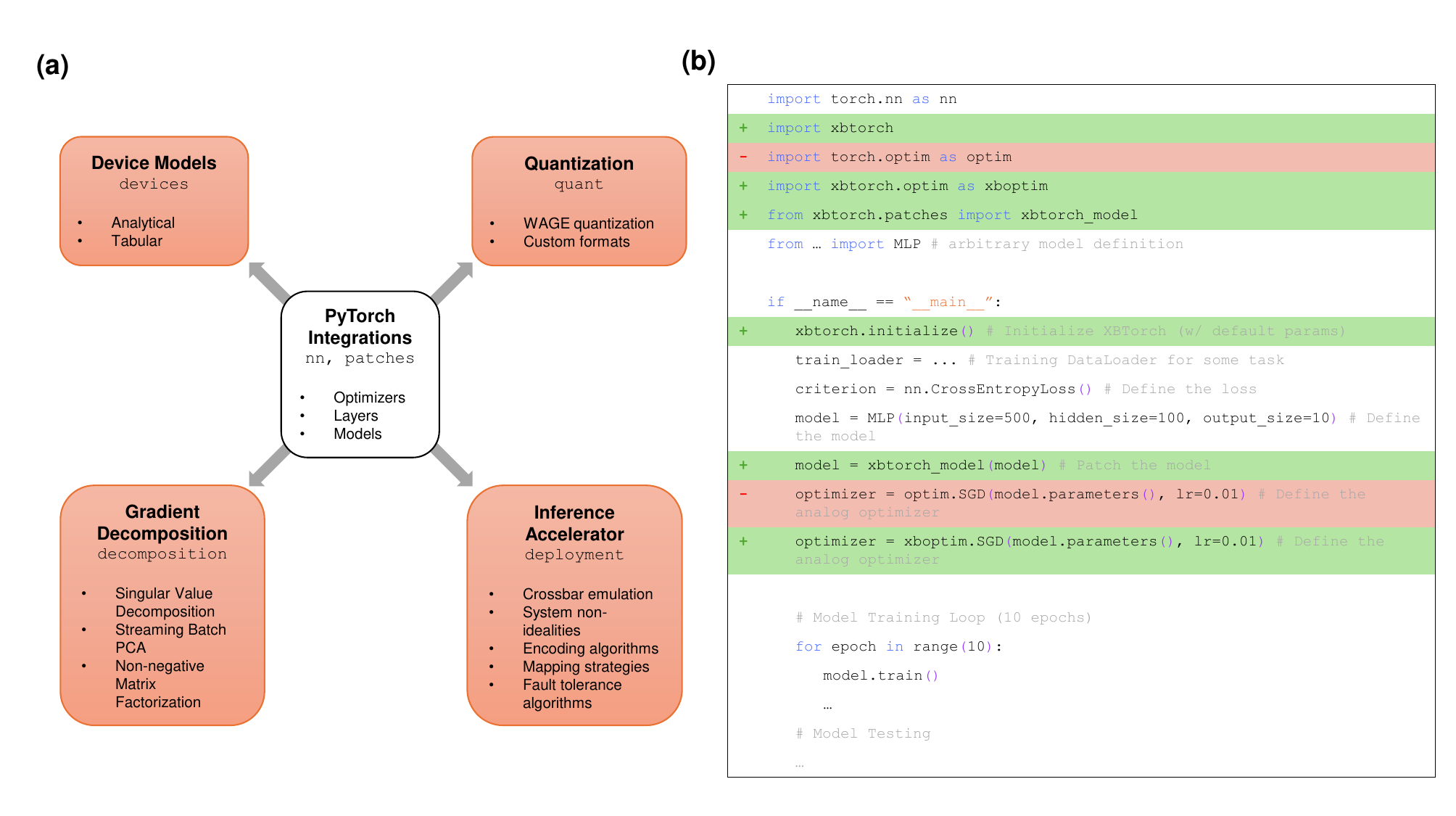}
  \caption{XBTorch features and an example script comparison. (a) XBTorch features (in \textbf{bold}), corresponding library modules (in \texttt{monospace}), and lists of important key capabilities. (b) A code diff for converting a PyTorch script for training a simple neural network to XBTorch. XBTorch requires minimal changes for initialization and model patching.} 
  \label{fig:code_comparison}
\end{figure*}

Fig. \ref{fig:code_comparison}a presents a high-level overview of key features in XBTorch. XBTorch is designed with the philosophy that an effective memristive neural network simulator should provide both accuracy in modeling hardware constraints and flexibility in general algorithm development. The codebase is highly modular and prioritizes uniformity with PyTorch counterparts across all operations as much as possible. The result is a familiar API which users experienced with PyTorch can quickly comprehend and experiment with, as shown in Fig. \ref{fig:code_comparison}b. Specific features and modules are detailed below.

\subsection{Device Models} 
\label{sec:devicemodel}

The \texttt{devices} module implements two types of device models: \textit{analytical} and \textit{tabular}. Analytical models are based on an explicit mathematical formulation that allows explicitly specifying device-level noise sources such as cycle-to-cycle variability, device-to-device variability, and non-linearity, while tabular models are based on pre-populated jump tables or lookup tables that implicitly capture such noise sources. We provide several \textit{presets} for both types. These are out-of-the-box models that are usable without any customization, but advanced users can still parameterize individual presets as needed. All presets are either based on experimental examples reported in literature, or synthesized based on some ground truth switching distribution. Examples of analytical model presets are the \textit{real} and \textit{ideal} device models for ReRAM devices from NeuroSim \citep{peng2019dnn+}, while examples of tabular model presets are the lookup table-based models for ReRAM as well as FeFET devices from previous works \citep{yousuf2023device, yousuf2025robust}. A tabular equivalent of the real analytical device preset is also provided to serve as a reference. We include a comparison between these two in terms of the total time to update neural network weights as a function of network depth in our public repository. 

The \texttt{devices} module also implements functions for creating custom device models. Given initial conductance, change in conductance per applied voltage pulse datasets, users can seamlessly create custom device models with arbitrary responses using state-of-the-art Kriging-based interpolation methodologies \citep{hossen2022data}. Custom models can be loaded and utilized as drop-in replacements of existing presets, allowing one to study the network training performance of custom devices as an example. We also include detailed examples on using as well as customizing device models in our public repository.

The process of weight updates with an XBTorch device model can be summarized as follows: each trainable parameter \(\theta_{ij}\) is converted to an equivalent device conductance \(G_{ij}\) and the corresponding backpropagated gradient of the loss function \(\nabla_\theta L_{ij}\) is converted to an integer number of pulses as follows:

\begin{equation}
    G_{ij} = \left( \frac{\theta_{ij} - W_{\min}}{W_{\max} - W_{\min}} \right) (G_{\max} - G_{\min}) + G_{\min} ,
    \label{eq:gToW}
\end{equation}

\begin{equation}
    p_{ij} = \boldsymbol{round} ( p_{max} \cdot \nabla_\theta L_{ij} ) ,
\end{equation}

where \(p_{max}\) is a scaling parameter defined as the total number of pulses required to fully set or reset the device and \(\boldsymbol{round(\cdot)}\)  represents stochastic rounding. The \(p_{ij}\) pulses are then applied to the device at initial conductance \(G_{ij}\) via the device model to update the conductance, which is then finally converted back to the weight domain \(\theta_{ij}\) for the next epoch (using the inverse operation of Eq. \ref{eq:gToW}).






\subsection{Hardware-aware Training}
\label{sec:hwatraining}

All training experiments that utilize an XBTorch device model are hardware-aware in the sense that weight updates are noisy, with the overall noise corresponding to the switching noise present in the underlying device being modeled. To model system-level non-idealities, these noisy updates can be paired with further noise injection techniques that are either explicit (such as injection of gaussian noise explicitly to network quantities) or implicit (such as quantization of network quantities, decomposing network gradients to low-rank approximations, etc.). XBTorch device models can be used in conjunction with such injection techniques to realize and study more advanced hardware-aware training schemes.

XBTorch provides a straightforward way of implementing such techniques, with out-of-the-box support for WAGE quantization \citep{wu2018training} and several gradient decomposition methods (covered in Section \ref{sec:decomposition}). WAGE quantization is a quantization-aware training paradigm where all network quantities i.e. weights (W), activations (A), gradients (G), and errors (E)  are discretized to low-bitwidth integers. From the perspective of memristive neural networks, the 2-8-8-8 WAGE configuration is particularly important as it produces ternary weight networks that are easily mappable to arrays of emerging memory devices (even when the tunability of devices is limited) and robust to system-level noise sources  \citep{yousuf2025robust}.

Primitives for activations and errors are handled by the \texttt{quant} module, while primitives for weights and gradients are handled by the \texttt{patches} module by wrapping over PyTorch optimizers. This separation is a natural result of how  forward and backward passes happen separately in PyTorch. XBTorch maintains a simple and scalable interface for users to implement their own quantization functions. It also integrates with QTorch \citep{zhang2019qpytorch}, as well as common large language model (LLM)-serving and evaluation tooling such as HuggingFace's \texttt{transformers} \citep{wolf2019huggingface} and EleutherAI's \texttt{lm-eval} harness \citep{eval-harness}, enabling the specification of completely custom number formats for all network quantities and facilitating research in quantization schemes for memristive neural networks as well as scalable, standardized benchmarking of large language models under analog noise sources.

In XBTorch, we choose to accurately simulate device-level noise sources (as explained in Section \ref{sec:devicemodel}) during weight updates via device models. This is similar to how device models behave in classic simulators like NeuroSim, but different to noise injection operations in modern simulators like AIHWKIT-lightning, which approximate weight update noise at a higher-level for faster simulation times when training. As a result, hardware-aware training results from XBTorch would be more representative of network performances that could be attained in hardware realizations, especially for small-scale neural network experiments that are not heavily over-parameterized. We improve simulation times by ensuring that device models during hardware-aware training are \textit{stateless} instead of \textit{stateful} (detailed further in Section \ref{sec:hwainference}).

\subsection{Gradient Decomposition Methods}
\label{sec:decomposition}

Training memristive neural networks demands substantial memory and computational resources, primarily because it involves generating and averaging large gradient matrices constructed via outer products over batches of training data. Low-rank approximations provide an effective strategy to ease this burden by compressing the gradient information, thereby reducing the overall data footprint. Given a weight matrix \(\theta\), its full-rank gradient for loss function \(L\) over batch \(B\) can be approximated as a matrix with a lower-rank $k$ as \(\nabla_{\theta}^{(k, B)} l\), such that:

\begin{equation}
    \nabla_{\theta}^{(k, B)} l = \Delta \cdot \Sigma \cdot X^T
    \label{eq:gradient_decomp} ,
\end{equation}

This reduction is particularly advantageous in distributed computing environments, where it can help decrease communication overhead and energy usage \citep{vogels2019powersgd}. It also has importance in the context of memristive neural networks as compressed gradients require fewer device writes, which can greatly benefit devices given their write limitations \citep{hoskins2019streaming}. Even in more traditional CMOS systems, such compression techniques can yield significant energy savings during the training process \citep{huang2023low}.

XBTorch facilitates the investigation of gradient decomposition methods by implementing easily extensible primitives via the \texttt{decomposition} module that abstract away network-level details (the user only needs to provide details of how to decompose gradient matrices, everything else is managed by XBTorch). XBTorch implements the Streaming Batch Principal Component Analysis (SBPCA) algorithm from \cite{huang2023low}, Non-negative Matrix Factorization (NMF), and Singular Value Decomposition (SVD)-based compression techniques for gradient decomposition.

If used in conjunction with hardware-aware training, the order of operation is as follows: error quantization, gradient quantization, gradient compression, conductance updates  via device models and compressed gradients. This operation order reflects a hardware architecture where analog arrays interface with a digital co-processor: current measurements are read via analog-to-digital converters (ADCs), low-rank gradient updates are computed by the co-processor, and updates are applied back to the arrays through digital-to-analog converters (DACs).

\subsection{Loss Landscapes}

Loss landscapes \citep{li2018visualizing} provide a way to visualize how a given neural network behaves when its parameters are perturbed. 
\begin{equation}
    f(\alpha, \beta) = L\left(\vec{\theta}_i + \alpha \frac{\vec{\delta}}{\|\vec{\delta} \|} + \beta \frac{\vec{\eta}}{\|\vec{\eta} \|}\right)
    \label{eq:loss} ,
\end{equation}
where $\vec{\theta}_i$ is simply a column vector corresponding to (unaltered) network parameters after training epoch \(i\), \(\vec{\delta}\) and \(\vec{\eta}\) represent direction vectors (having similar dimensions as \(\vec{\theta}_i\)) that can be chosen either randomly or via some structured methodology, and \(\alpha\), \(\beta\) are scaling parameters that dictate the degree of the overall perturbation applied to the network. For some value of \(\alpha\) and \(\beta\), the perturbation is applied to network parameters and then the loss function \(L\) is computed over the entire training set. To visualize the full loss landscape, the quantity in Eq. \ref{eq:loss} can be computed over a grid of values of \(\alpha\) and \(\beta\).

Loss landscapes can also be used to visualize parameter optimization trajectories. However, since neural network optimization trajectories are known to lie in extremely low-dimensional spaces, meaningful direction vectors \(\vec{\delta}\) and \(\vec{\eta}\) are required. The authors in \citep{li2018visualizing} demonstrate that optimization trajectories can be effectively utilized if one sets \(\vec{\delta}\) and \(\vec{\eta}\) to be the first two principal components of the matrix of parameter evolutions given by
\begin{equation}
    \boldsymbol{M} = [\vec{\theta}_0 - \vec{\theta}_n, \dots, \vec{\theta}_{n-1} - \vec{\theta}_n] .
\end{equation}
Once the principal components are known, \(\boldsymbol{M}\) can be projected to the sub-space spanned by them via a simple dot product operation, giving a direct and effective mechanism to visualize the optimization trajectory of network parameters.

Visualizing loss landscapes and optimization trajectories in this manner provides valuable insights into the behavior and performance of memristive neural networks. Non-idealities inherent to these systems - such as device-level drift, read noise, and programming variability - can be effectively modeled as parameter perturbations, analogous to those described in Eq. \ref{eq:loss}. Additionally, such visualizations offer insight into the effective dimensionality of network optimization, making them a useful tool for analyzing device-network interactions. They also enable comparative evaluation of different device models or hardware-aware training strategies, particularly in terms of robustness to noise and parameter perturbations.

The XBTorch \texttt{loss} module provides tools that simplify computing and visualizing loss landscapes and parameter optimization trajectories for memristive neural networks. A detailed example notebook is also included in the public repository.

\subsection{Hardware-aware Inference}
\label{sec:hwainference}

The \texttt{deployment} module handles everything related to hardware-aware inference. This module is intended for use cases where one is interested in gauging what the inference performance of a pre-trained network (with or without hardware-aware training) would be if deployed to a realistic hardware system with device-level as well as system-level noise sources. The module maintains a stateful representation of underlying crossbar arrays, which we term as the underlying \textit{inference accelerator}. A stateful representation means that the inference accelerator maintains the complete crossbar array at all times. The accelerator is parameterized with dimensions of the crossbar arrays. If a pre-trained network has to be mapped to an accelerator for hardware-aware inference, the user must also specify a \textit{weight encoding} algorithm, which describes how software weights are encoded into device conductances, a \textit{conductance mapping} algorithm, which dictates where should encoded conductances be mapped on the accelerator, and an \textit{output polling} parameter, which specifies how output currents from (possibly multiple) mapped instances of layers should be gathered. Because of this representation, it is possible to run out of devices if there aren't enough devices in the accelerator to map a particular neural network solution. This stateful representation is contrary to the stateless representation in the device modeling feature of XBTorch described in Section \ref{sec:devicemodel}, where devices are assumed to be always available regardless of the number of network parameters.

The stateful representation enables investigations of mapping algorithms, encoding algorithms, as well as combinations of the two for hardware-oriented fault-tolerance algorithms. Most simulators assume a stateless representation in hardware-aware inference. Although that approach has benefits in terms of simulation times, it is fundamentally limited as it does not allow investigations into inference-time fault-tolerance algorithms. Research in such algorithms has been somewhat disjoint compared to research in building simulators of memristive neural networks for training purposes.  By maintaining a stateful representation during hardware-aware inference, XBTorch provides a unified interface that can effectively bridge these domains. XBTorch natively implements the mapping algorithm with inner fault tolerance \citep{huangfu2017computation}, the layer ensemble averaging algorithm \citep{yousuf2025layer}, and the committee machines algorithm \citep{joksas2020committee} for exploratory or comparative studies.

In addition to the stateful crossbar emulation used for layer-level mapping and inference, XBTorch also supports hardware-aware evaluation of large language models (LLMs). Because LLM parameter counts commonly run into the billions and maintaining a full stateful accelerator instance for each parameter is prohibitively memory intensive, hardware-aware LLM evaluation in XBTorch is implemented in a stateless mode: instead of instantiating a persistent crossbar allocation for every model parameter, XBTorch streams weight-to-conductance encodings and performs per-inference emulated reads and noisy MAC computations on demand. This stateless approach reduces peak memory usage while preserving key analog non-idealities (read/write noise, ADC/DAC quantization, conductance overlap, etc.) in the inference loop, enabling realistic evaluation of LLM behavior on crossbar accelerators at manageable resource cost. To ease adoption for standard benchmarking workflows, XBTorch provides adapters to common LLM-serving and evaluation tooling (e.g., HuggingFace's \texttt{transformers}\citep{wolf2019huggingface} for versatile, memory-aware inference and EleutherAI's \texttt{lm-eval} harness\citep{eval-harness} for standardized task evaluation).

Like the \texttt{device} module, we provide accelerator presets with detailed examples on hardware-aware inference in the public repository, ensuring extensibility towards custom platforms for users and device researchers. We also include a comparative example on the aforementioend inference-time fault tolerance schemes.

\section{Results}

To illustrate the capabilities of XBTorch and the kinds of algorithmic investigations one can carry out, we present case studies using a simple multi-layer perceptron network for classifying images of handwritten digits from the MNIST dataset as a running example. Further details of this network architecture and all algorithmic hyperparameters are presented in Appendix \ref{app:hyperparams}. The public repository contains detailed example notebooks and guides for recreating all reported results. 

\subsection{Hardware-Aware Training} 

\begin{figure*}[htp]
  \centering
  \includegraphics[width=0.8\linewidth]{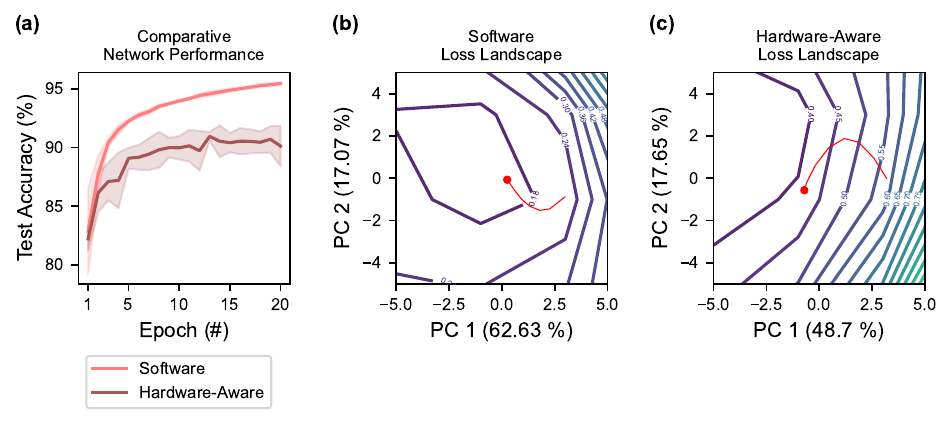}
  \caption{Comparison of a multi-layer perceptron network trained on MNIST using a low-precision hardware-aware training in XBTorch against a full precision software baseline trained using vanilla PyTorch. (a) Network test accuracies averaged over 10 independent runs. Loss landscapes for the (b) software baseline network and the (c) hardware-aware network. Loss landscape axes correspond to principal components over the parameter evolution matrix \(\boldsymbol{M}\), with explained variances presented in parentheses. The optimization trajectories are included as red lines concluding at the circle marker. The hardware-aware network is trained using a 2-8-8-8 WAGE configuration which produces ternary weight networks, and a tabular FeFET device model with \(V_{gs} = 0.9\) \(V\) and \(1 \%\) variability (see Fig. \ref{fig:tabular_models}).
  } 
  \label{fig:hwa_training}
\end{figure*}

The hardware-aware training feature in XBTorch can be used to compare the training performance of vanilla software networks against device networks, as well as compare two device networks (trained with differing device models) against one another. Loss landscape visualizations can also aid in these comparisons, as they can be used as a tool to assess either the robustness of different hardware-aware training schemes or device-network interactions in general.

In Fig. \ref{fig:hwa_training}, we compare hardware-aware training results of a software baseline network with a device network using a tabular FeFET model with low variability, and WAGE quantization with the 2-8-8-8 setting.  This hardware-aware training scheme is similar to the scheme described in \citep{yousuf2025robust}, and produces robust ternary weight networks that can be easily deployed to emerging memory device arrays even with strict restrictions on device tunability. The full list of hyperparameters is presented in Appendix \ref{app:hyperparams} 

It can be seen that the performance of software networks is consistently higher than the hardware-aware networks. This is a direct result of the degree of implicit and explicit noise injection operations that are happening as a result of the hardware-aware training paradigm (noisy weight updates, quantized network quantities). These noise sources essentially act as a regularizing mechanism. Corresponding loss landscapes were computed by perturbing only the first fully connected layer, since it was the largest and thus likely accumulates more information during training compared to following layers. All sources of randomness were fixed, except for the underlying device model. An interesting point to note here is that compared to the software baseline, trajectories of the hardware-aware network lie in higher dimensional spaces, as the two principal components only explain \(\approx 60 \%\) of the variance in the hardware-aware case compared to \(80 \%\) in software. This hints to the fact that device and system-level sources of noise inject noise in backpropagated gradients, making device networks more difficult to train. The optimization trajectories show that the two networks converge to drastically different local minima. Since all sources of randomness apart from the underlying device model were fixed, we conclude that this is a direct result of device-level switching behavior.

\subsection{Gradient Decomposition Methods}
\label{res:decomposition}

\begin{figure*}[htp]
  \centering
  \includegraphics[width=0.9\linewidth]{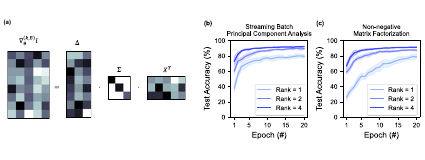}
  \caption{Demonstration of gradient decomposition methods using XBTorch. (a) Representative decomposition of a hypothetical gradient matrix into a low-rank \(k\) representation according to Eq. \ref{eq:gradient_decomp}. The decomposed form has less memory overhead compared to the full rank gradient matrix if \(k\) is small. Network test accuracies averaged over 10 independent runs at different ranks for (b) SBPCA and (c) NMF algorithms. 
  } 
  \label{fig:decomposition}
\end{figure*}

Fig. \ref{fig:decomposition}a depicts a visualization of a sample low-rank gradient decomposition into its left and right singular vector matrices, as described previously in Eq. \ref{eq:gradient_decomp}. For an \(m \times n\) gradient, a rank-\(k\) approximation requires storing \(O(k(m+n))\) elements. For large matrices, one can thus extract substantial memory savings when \(k << min(m, n)\).

We re-train the same network architecture with firstly SBPCA and then NMF at increasing ranks. We choose a higher variability FeFET device model without WAGE quantization compared to Fig. \ref{fig:hwa_training}. These choices are solely for the purpose of showing the variety of options that XBTorch provides. The full list of hyperparameters is presented in Appendix \ref{app:hyperparams}. Network training results for the two algorithms are summarized in Fig. \ref{fig:decomposition}b and \ref{fig:decomposition}c respectively. 

We see that the performance of the two algorithms is quite different, especially at low-ranks. This is because the decompositions that they produce exhibit unique properties, causing differences in device-network interactions. SBPCA has an oscillatory nature since it can produce negative-valued decompositions even for strictly positive gradient matrices. In contrast, NMF always produces additive features because of non-negative constraints. At ranks 1 and 2, SBPCA outperforms NMF as it can train networks to a higher accuracy at similar training times, though their performance becomes similar at ranks at and beyond 4. Despite this gap, one may still choose to use NMF over SBPCA because the additive nature reduces overlaps between rank-by-rank updates, which may be beneficial for non-linear, retention-limited devices where oscillatory updates are highly detrimental. XBTorch introduces a unified interface for studying how such decomposition algorithms interplay with device characteristics and neural network performance.

\subsection{Hardware-aware Inference \& Fault-Tolerance Schemes}

XBTorch allows users to define the overall dimensions of the simulated crossbar when performing hardware-aware inference. System-level parameters that introduce noise - such as ADC/DAC bit precision or stuck percentage - can also be specified. Based on these settings, XBTorch simulates a stateful crossbar representation using the accelerator primitive described in Section \ref{sec:hwainference}. Pre-trained solutions can then be written to an accelerator instance using chosen weight (weight-to-conductance) encoding and crossbar (conductance-to-device) mapping algorithms.

To showcase these features, we demonstrate hardware-aware inference by using a hardware-aware trained solution from Fig. \ref{fig:hwa_training} on a simulated crossbar of size \(2500 \times 2500\). We use a differential weight encoding scheme, which encodes each weight matrix \(\theta\) to two conductance matrices, \(G_{pos}\) and \(G_{neg}\), such that \(\theta \propto (G_{pos} - G_{neg})\), along with a random crossbar mapping algorithm, which maps the conductance matrices to random non-conflicting positions on the simulated crossbar. When devices are being tuned, write noise is simulated; users can choose between various distributions to match desired device physics. Finally, during inference, normalized dataset inputs are converted to voltages and scaled by a specified \(V_{read}\) voltage. Then, the voltages are applied to the input lines of the two differentially mapped conductance matrices by performing an explicit noisy readout and performing vector-matrix multiplications between the input voltages and the device conductances. This generates differential output currents corresponding to the differentially mapped conductances. Like the writing phase, read noise is also simulated; users can specify the noise level to simulate noisy ADC/DAC readouts. From the differential currents, the final output is assembled by computing their difference, followed by appropriate scaling. This operation, along with any non-linear activation function, does not include explicit noise sources. XBTorch thus makes an assumption that such operations happen in a digital co-processor that is separate from the analog accelerator comprising of memristive crossbars. The full list of hyperparameters is presented in Appendix \ref{app:hyperparams}

Fig. \ref{fig:hwa_inference}a shows a readout of device conductances, or a conductance map, after the write operation. The two layers are marked by separate colors, and a zoomed-in block is shown as an inset for clarity. Fig. \ref{fig:hwa_inference}b shows the probability distribution of devices in \(G_{pos}\) of the first network layer.  The overlap between conductance states is a direct consequence of simulated noise sources during read and write operations from/to the accelerator. Finally, Fig. \ref{fig:hwa_inference}c showcases the hardware-aware inference performance obtained for this pre-trained solution as a function of ADC/DAC bit precision (keeping all other noise sources fixed). As expected, the inference performance approaches improves with increasing bit precision. We include a baseline accuracy estimate for reference; this is the hardware-aware inference performance on a fully ideal accelerator (i.e. no noise sources). Even at the highest level of ADC/DAC bit precision, the hardware-aware inference performance lags behind this ideal baseline because of the presence of other noise sources that result in the conductance overlap visible in \ref{fig:hwa_inference}b. For reference, a list of hyperparameters used for this experiment is presented in Appendix \ref{app:hyperparams}.

\begin{figure*}[htp]
  \centering
  \includegraphics[width=0.9\linewidth]{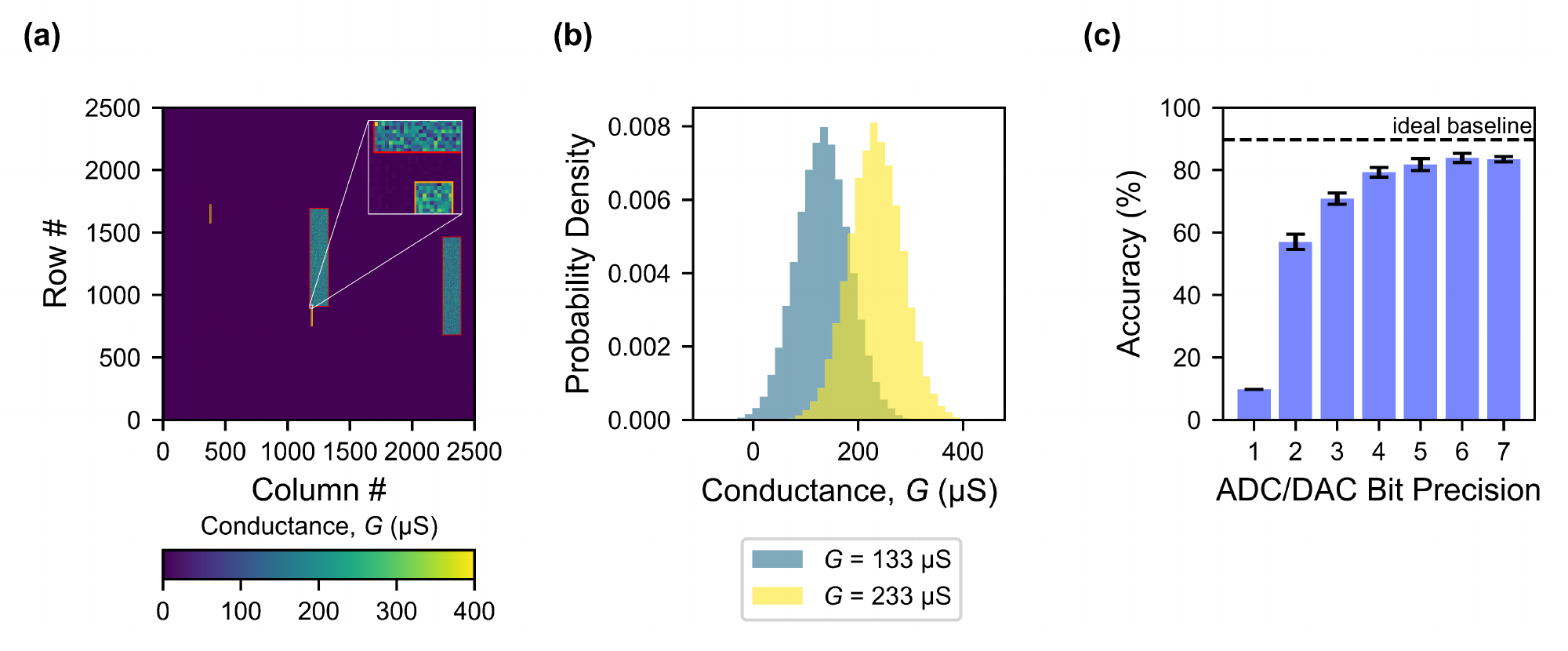}
  \caption{Hardware-aware inference using XBTorch. (a) Simulated \(2500 \times 2500\) crossbar with realistic noise shows conductance maps after writing a trained solution (from Fig. \ref{fig:hwa_training}). Each weight matrix \(\theta\) is differentially encoded into \(G_{pos}\) and \(G_{neg}\), mapped to random, non-overlapping locations. Red/orange boxes mark the first/second network layers; dark/light shades represent \(G_{pos}\)/\(G_{neg}\). Insets show a \(25 \times 25\) zoomed region. Unused devices are disabled during readout. (b) Actual conductance distributions for \(G_{pos}\) from the first layer. (c) Inference-time test accuracy as a function of ADC/DAC bit precision. The dotted line corresponds to the ideal baseline with no noise sources. Bar heights are averages across 10 independent iterations. Error bars correspond to 3 standard deviations.
  } 
  \label{fig:hwa_inference}
\end{figure*}

\begin{figure*}[htp]
  \centering
  \includegraphics[width=0.7\linewidth]{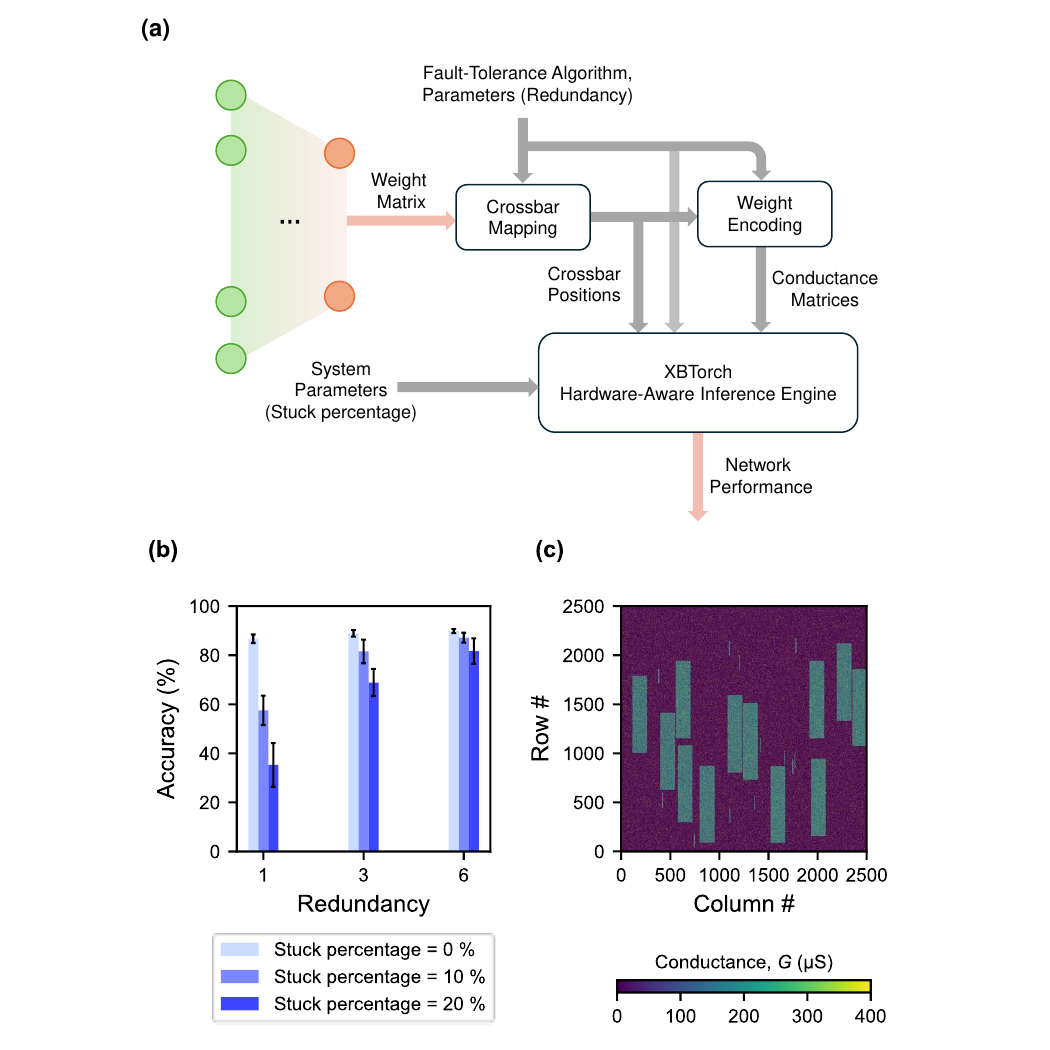}
  \caption{Demonstration of layer ensemble averaging in XBTorch. (a) The overall flow of mapping an individual network layer to an emulated crossbar using XBTorch. System parameters, such as total crossbar size and a percentage of stuck devices within that crossbar can be specified to gauge network performance under realistic noise sources. (b) Network inference accuracy as a function of mapping redundancy and percentage of stuck devices. Under layer ensemble averaging, outputs are averaged layer-wise from the redundant mappings during inference. Bar heights correspond to averages across 10 independent iterations, and error bars correspond to a single standard deviation. (c) Conductance map after encoding and mapping the network solution for the case where the redundancy level is \(6\) and \(20 \%\) devices are stuck.
  } 
  \label{fig:fault_tolerance}
\end{figure*}

Fault-tolerance algorithms are another key feature in XBTorch. An effective fault-tolerance algorithm can increase the inference performance of a memristive neural network in the presence of even severe non-idealities - both at the device-level (such as stuck-at faults) and system-level (such as ADC/DAC limitations).  The stateful representation and the separation between encoding and mapping phases in XBTorch enables one to seamlessly specify a fault-tolerance algorithm on top of any existing hardware-aware inference application. Fig. \ref{fig:fault_tolerance}a demonstrates how various components flow into the XBTorch hardware-aware inference engine when using a fault-tolerance algorithm. One can either choose from a pre-implemented fault-tolerance algorithm, or easily implement and utilize a customized algorithm thanks to our streamlined preset design methodology.

To demonstrate this feature, we showcase results using layer ensemble averaging (LEA) \citep{yousuf2025layer} as the fault-tolerance algorithm. In LEA, one has to specify a redundancy parameter, which dictates how many times each encoded conductance matrix gets mapped on the accelerator crossbar. During inference, currents induced by each of the two differentially encoded conductance matrices are averaged across the redundant mappings for each layer. These averaged currents are then used for following computations. The averaging has the net effect of decreasing the mapping error of neural network layers onto noisy device crossbars, which in turn can increase the overall performance of the network.

Fig. \ref{fig:fault_tolerance}b shows the hardware-aware inference performance of LEA as a function of increasing redundancy (number of times each conductance matrix is mapped) and stuck percentage (the percentage of devices out of \(2500 \times 2500\) that are either stuck-high or stuck-low). At any given redundancy level, a higher percentage of stuck-at faults decreases the network's inference performance. This is expected, as more stuck devices increase the network mapping error directly. The fault-tolerance algorithm shines when we examine increasing levels of redundancy at a fixed level of stuck-at faults. In the worst case of stuck-at faults, i.e. \(20\%\), LEA can improve the network's performance from \(< 40\%\) to \(> 80\%\), which is within \(5\%\) of the case where there are no stuck-at faults. Fig. \ref{fig:fault_tolerance}c showcases a conductance map similar to Fig. \ref{fig:hwa_inference}a for the case with a redundancy level of \(6\) and a stuck percentage of \(20\%\). As can be seen, all redundant mappings of each layer are represented in a stateful fashion to accurately simulate memristive neural network inference on realistic systems, which we believe is an important precondition for investigations into inference-time fault-tolerance algorithms in general.    

\subsection{Language Model Evaluation}

\begin{table*}[htp]
\centering
\small
\setlength{\tabcolsep}{5pt}
\renewcommand{\arraystretch}{1.2}
\begin{tabular}{llccccccc}
\toprule
\textbf{Model} & \textbf{ADC/DAC Bits} & 
\makecell{\textbf{Arc-Easy} \\ (10-shot)} & 
\makecell{\textbf{Arc-Challenge} \\ (10-shot)} & 
\makecell{\textbf{BoolQ} \\ (0-shot)} & 
\makecell{\textbf{PIQA} \\ (0-shot)} & 
\makecell{\textbf{Winogrande} \\ (5-shot)} & 
\textbf{Avg.} $\uparrow$ & \textbf{Stderr} \\
\midrule

\multirow{4}{*}{\textbf{TriLM-830M}} 
 & 4  & 25.93 & 21.84  & 41.16 & 53.21 & 53.20 & 39.07 & 0.50 \\
 & 8  & 53.87 & 22.95  & 61.59 & 65.94 & 53.04 & 51.48 & 0.51 \\
 & 12 & 56.31 & 24.49  & 60.58 & 68.23 & 53.75 & 52.67 & 0.51 \\
 & 16 & 56.23 & 24.57  & 60.49 & 68.23 & 54.06 & 52.72 & 0.51 \\
 \midrule
\multirow{4}{*}{\textbf{TriLM-3.9B}} 
 & 4  & 25.97 & 22.61 & 42.26 & 51.14 & 52.33 & 38.86 & 0.51 \\
 & 8  & 61.91 & 30.12 & 61.68 & 67.79 & 57.06 & 55.71 & 0.52 \\
 & 12 & 68.18 & 34.98  & 67.61 & 73.83 & 64.96 & 61.91 & 0.50 \\
 & 16 & 67.80 & 34.90 & 67.52 & 73.94 & 65.04 & 61.84 & 0.50 \\

\bottomrule
\end{tabular}
\caption{
Performance of \textbf{TriLM} models under varying ADC/DAC bit precisions using XBTorch inference accelerators. Reported numbers are accuracy (\%) across multiple benchmarks, with the last two columns showing mean performance and its standard error.
}
\label{tab:trilm_bits_precision}
\end{table*}

We evaluated the hardware-aware, stateless LLM inference flow on a set of standard natural-language benchmarks. Table \ref{tab:trilm_bits_precision} reports task-level accuracies for two models of varying sizes from the TriLM model family \citep{kaushal2024spectra} (high-performing ternary weight networks) across varying ADC/DAC fixed-point precisions. All accelerator and inference run parameters used to generate these results (device conductance ranges, read/write noise levels, read voltage, etc.) are reported in Appendix \ref{app:hyperparams}. The experiments demonstrate the sensitivity of LLM task performance to ADC/DAC precision and highlight how analog non-idealities captured by XBTorch could translate into end-task degradations and the potential headroom for fault-tolerance or higher-bit ADC/DAC hardware. It is also interesting to see how larger model sizes are inherently more robust to analog noise compared to smaller model sizes across all evaluated tasks. Because of the stateless representation and the native integration with \texttt{transformers} and \texttt{lm-eval}, researchers can easily swap out the models, evaluation tasks, as well as underlying accelerator parameters with custom measurements to obtain accurate performance estimates in a memory-efficient fashion. We provide brief descriptions of the evaluation tasks used in Table \ref{tab:trilm_bits_precision} in Appendix \ref{app:lmdatasets}.

\section{Conclusion}

We introduced XBTorch, a framework designed to streamline algorithmic research on memristive neural networks. XBTorch aims to bridge existing gaps in the field and foster broader collaboration. While the framework offers a cohesive set of features, some limitations remain - particularly for hardware-aware large model training. To address this, we are actively working on fast noise approximations for training purposes as a first step. We envision XBTorch as a community-driven effort to advance algorithmic investigations in memristive neural networks and the broader field of neuromorphic computing in general.

\begin{acknowledgments}
 The work of Osama Yousuf was sponsored by the Army Research Laboratory and was accomplished under
Cooperative Agreement Number W911NF-25-2-0073. The views and conclusions contained in this
document are those of the authors and should not be interpreted as representing the official
policies, either expressed or implied, of the Army Research Laboratory or the U.S. Government.
The U.S. Government is authorized to reproduce and distribute reprints for Government purposes
notwithstanding any copyright notation herein.
\end{acknowledgments}

\section*{Data Availability Statement}

All data related to the findings of this study can be generated via examples available at the XBTorch repository: \url{https://github.com/ADAM-Lab-GW/xbtorch}.

\appendix

\counterwithin{figure}{section}
\counterwithin{table}{section}

\section{Device Model Presets}
\label{app:dev_model_presets}

\begin{figure*}[htp]
  \centering
  \includegraphics[height=0.5\linewidth]{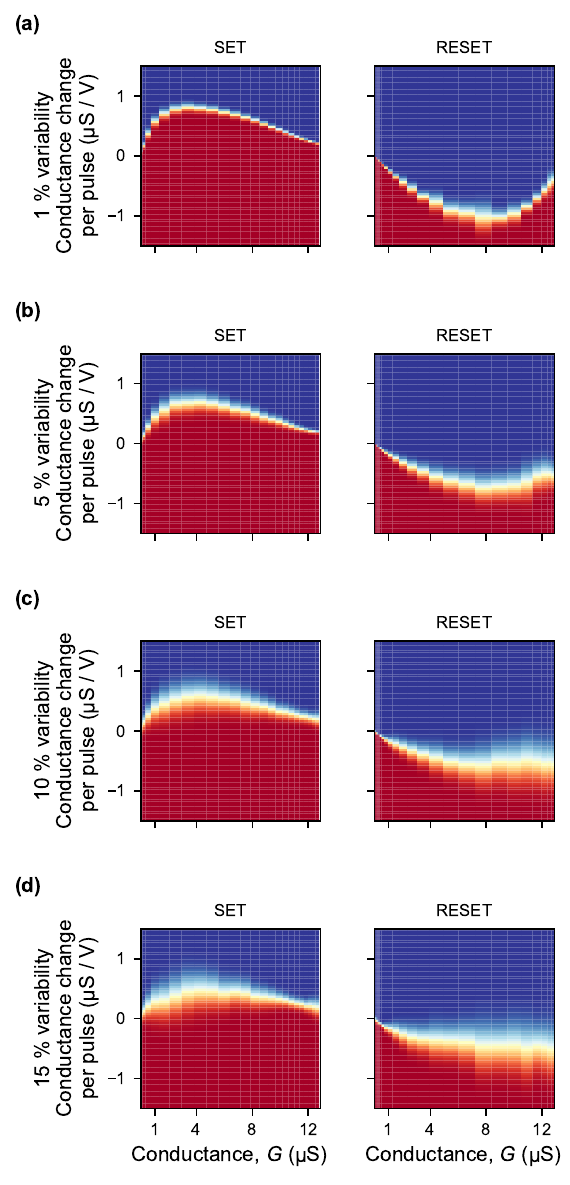}
    \caption{Cumulative Distribution Function (CDF) arrays for the FeFET tabular device model presets at \(V_{gs} = 0.9\) \(V\) and increasing levels of variability.}
  \label{fig:tabular_models}
\end{figure*}

XBTorch provides a wide variety of device model presets. For example, when using tabular models for FeFET devices, one can specify a wide range of device-to-device variability levels and gate-source voltage \(V_{gs}\) values as parameters. As these parameters change, the overall switching noise of the devices changes, leading to differences in neural network-level performance. Fig. \ref{fig:tabular_models} shows the tabular arrays for the SET as well as RESET phases for the available FeFET device models at \(V_{gs} = 0.9\) \(V\) and increasing levels of variability.

\section{Hyperparameters}
\label{app:hyperparams}
For all non-language model results, we used a simple 2-layer perceptron network with 784 input neurons, 150 hidden neurons, and 10 output neurons. We used cross entropy as the loss function. The input dataset is transformed to be zero mean and unit variance. When training, the data is presented to the network in randomly shuffled mini-batches of size 4096. Hyperparameters for individual experiments are reported in the tables below.

\begin{table*}[htp]
\centering
\begin{tabular}{|c|c|}
\hline
\textbf{Hyperparameter} & \textbf{Value} \\
\hline
Learning rate (software) & 0.1 \\
Device model & Tabular FeFET: \(V_{gs} = 0.9\) \(V\), var = 1\% \\
Learning rate (hardware-aware) & 4.76 \\
WAGE bit-widths & 2, 8, 8, 8 \\
WAGE rounding & nearest \\
Weight range & (-1, +1) \\
Optimizer & \texttt{xboptim.SGD} \\
\hline
\end{tabular}
\caption{Hyperparameters for results in Fig. \ref{fig:hwa_training}.}
\end{table*}

\begin{table*}[htp]
\centering
\begin{tabular}{|c|c|}
\hline
\textbf{Hyperparameter} & \textbf{Value} \\
\hline
Learning rate (all ranks) & 0.1 \\
Device model & Tabular FeFET: \(V_{gs} = 0.9\) \(V\), 5\% variability \\
Weight range & (-1, +1) \\
Optimizer & \texttt{xboptim.SGD} \\
\hline
\end{tabular}
\caption{Hyperparameters for results in Fig. \ref{fig:decomposition}.}
\end{table*}

\begin{table*}[htp]
\centering
\begin{tabular}{|c|c|c|}
\hline
\textbf{Hyperparameter} & \textbf{Value} \\
\hline
Number of devices supported by the accelerator (rows, columns) & (2500, 2500) \\
Minimum device conductance (\(\mu\)S) & 133 \\
Maximum device conductance (\(\mu\)S) & 233 \\
Read voltage (V) & 0.3 \\
Uniform read noise injected into the accelerator (\(\mu\)S) & 10 \\
Zero-mean Gaussian write noise (\(\mu\)S) & 50 \\
Scheme for encoding weights into device conductances & differential \\
Mapping scheme for device conductance matrices to accelerator positions  & random \\
 ADC/DAC precision (fixed-point precision of inputs/outputs to/from mapped layers) & 8 \\
Stuck percentage (\(\%\)) & 0.0 \\
Fault-tolerance algorithm & None \\
Scheme for aggregating outputs when layers are redundantly mapped & N/A \\
\hline
\end{tabular}
\label{table:hwainference}
\caption{Hyperparameters for results in Fig. \ref{fig:hwa_inference}.}
\end{table*}

\begin{table*}[htp]
\centering
\begin{tabular}{|c|c|c|}
\hline
\textbf{Hyperparameter} & \textbf{Value} \\
\hline
 ADC/DAC precision (fixed-point precision) & variable \\
Stuck percentage (\(\%\)) & variable \\
Fault-tolerance algorithm & LEA \\
Scheme for aggregating outputs when layers are redundantly mapped & average \\
\hline
\end{tabular}
\caption{Hyperparameters for results in Fig. \ref{fig:fault_tolerance}. Other parameters are kept the same as Fig. \ref{fig:hwa_inference}.}
\end{table*}

\begin{table*}[htp]
\centering
\begin{tabular}{|c|c|c|}
\hline
\textbf{Hyperparameter} & \textbf{Value} \\
\hline
Number of devices supported by the accelerator (rows, columns) & $(\infty, \infty)$ \\
Uniform read noise injected into the accelerator (\(\mu\)S) & 0 \\
Zero-mean Gaussian write noise (\(\mu\)S) & 0 \\
 ADC/DAC precision (fixed-point precision) & variable \\
Stuck percentage (\(\%\)) & - \\
Fault-tolerance algorithm & - \\
\hline
\end{tabular}
\caption{Hyperparameters for results in Table \ref{tab:trilm_bits_precision}. Other parameters are kept the same as Fig. \ref{fig:hwa_inference}. The accelerator is operated in stateless mode.} 
\end{table*}

\section{Language Model Evaluation Datasets}
\label{app:lmdatasets}

The following tasks were used to construct the language-model evaluation reported in Table \ref{tab:trilm_bits_precision}. For each, we used the open-source implementation available via \texttt{lm-eval}.

\subsection{ARC (AI2 Reasoning Challenge) — ARC-Easy / ARC-Challenge} 
ARC AI2 \citep{clark2018think} is a collection of multiple-choice, grade-school science questions; the dataset is partitioned into an Easy set and a Challenge set (the Challenge set contains questions that simple retrieval and word-cooccurrence methods fail to answer). We evaluate multiple-choice accuracy on both the ARC-Easy and ARC-Challenge subsets as provided by the standard task splits. 

\subsection{BoolQ} 
BoolQ \citep{clark2019boolq} is a naturally occurring yes/no question answering task where each example includes a question and a supporting passage; the model must predict a boolean (yes/no) answer. We evaluate BoolQ as a binary classification accuracy task using the canonical BoolQ splits. 

\subsection{PIQA (Physical Interaction: Question Answering)}
PIQA \citep{bisk2020piqa} is a two-choice multiple-choice dataset testing physical commonsense reasoning (choices correspond to plausible physical interactions or solutions). We measure selection accuracy on the PIQA test items. 

\subsection{WinoGrande} 
WinoGrande \citep{sakaguchi2021winogrande} is a large Winograd-schema style pronoun-resolution benchmark designed to measure commonsense reasoning while reducing dataset biases; examples are evaluated as a two-way choice (which candidate fills the pronoun). We report accuracy on the WinoGrande evaluation split.

\section*{References}   
\bibliography{references}

\end{document}